\documentclass[a4paper,10pt]{article}
\usepackage[top=0.9in,bottom=0.9in,left=0.75in,right=0.75in]{geometry}
\usepackage [dvips]{graphicx}
\usepackage{amssymb,color,amsmath,fancyhdr,setspace,caption}

 \title{Localization of a microtubule organizing center\\ by kinesin motors}
 \def\shorttitle{Localization of a MTOC by kinesin motors}
 \author{Chikashi Arita$^1$, Jonas Bosche$^1$, Alexander L\"uck$^2$, Ludger Santen$^1$} 
\def\shortauthor{C Arita, J Bosche, A L\"uck, L Santen} 
 \def\address{1 Theoretische Physik, Universit\"at des Saarlandes, 66041 Saarbr\"ucken, Germany\\
2 Informatik, Universit\"at des Saarlandes, 66041 Saarbr\"ucken, Germany}
 \date{\today}
 \def\abst{The microtubule (MT) motor Kip3p is very processive kinesin that promotes catastrophes and pausing in particular on cortical contact. These properties explain the role of Kip3p in positioning the mitotic spindle in budding yeast and potentially other processes controlled by kinesin-8 family members. We present a theoretical approach to positioning of a MT network in a cell. In order to explore the underlying mechanism we introduce an idealized system of two MTs connected by a microtubule organizing center (MTOC). The dynamics of Kip3p is modeled by interacting stochastic particles, which allows us to study the effects of motor-induced depolymerization under spatial confinement. We find that localization in the middle of the system is realized in a parameter regime where the motor densities on the MTs are increasing with the distance from the MTOC. Localization at an asymmetric position is also possible by tuning the kinesin input rates at the MT minus ends or attachment rates depending on different compartments of the cell. 
}

\makeatletter
\def\@maketitle{ 
\begin{center} 
 \let \footnote \thanks
 {\LARGE\linespread{1.2}\selectfont\textbf{\@title}\par} \vskip 10mm 
 {\Large \@author} \vskip 5mm 
 {\address} \vskip 5mm 
 \textbf{Abstract} \end{center}
 \begin{quote} \abst \end{quote} \vskip 5mm 
\noindent\makebox[\linewidth]{\rule{\textwidth}{0.5pt}}}
\makeatother

\fancypagestyle{titlepage}{
 \lhead{} \chead{} \rhead{}
 \lfoot{} \cfoot{\thepage} \rfoot{} }

\begin{document}

\maketitle
\thispagestyle{titlepage}

\maketitle

\section{Introduction} 
The cytoskeleton, a dynamic network of biopolymers, determines the shape of an eukaryotic cell. Microtubules (MTs), as well as actin and intermediate filaments, form the cytoskeleton. MTs consist of dimers, each of which is a set of $\alpha$ and $\beta$ tubulin subunits, and form cylindrical structures with high bending rigidity \cite{bib:Alb}. The cytoskeleton also plays the role of a transport network, and one important feature of MTs is the directionality; i.e., the two ends of a MT are distinguishable, as plus and minus ends \cite{bib:Kieke}.

Dynamic processes in a cell control the length distributions of MTs, so that MTs are adapted to different cell types and shapes. The regulatory mechanisms of MT lengths are carried out by polymerases and depolymerases. While polymerases support MT growth, depolymerases can induce fast depolymerization events, so-called catastrophes \cite{bib:GZH}. 
The length distributions are strongly influenced by the stabilization of growing MTs, in particular, under spatial confinement \cite{bib:ES}. Dynamically stabilized MTs connect the cell center and membrane, and keep their ability to adapt to different cell shapes. It has been also known that other proteins are able to enhance the depolymerization of MTs, which is the case for Kip3p, a member of the kinesin-8 family with extremely high processivity. Kip3p moves to the plus end of MTs and is known to promote catastrophes and pausing, and inhibited MT growth, most dramatically in contact with the cell cortex. The activity of Kip3p is of great importance in order to position the mitotic spindle in yeast \cite{bib:GCRP,bib:VHTHTH,bib:VLBDH}. 
 
The length regulation mechanism induced by Kip3p has been addressed in \cite{bib:HSMGMB,bib:RMF,bib:JEK,bib:MRF}.
In particular the models \cite{bib:RMF,bib:JEK,bib:MRF} capture the stepwise directed motion of kinesins and describe their interaction by mutual exclusion, which may be regarded as variants of the exclusion process \cite{bib:MGP,bib:Spitzer} with varying system size \cite{bib:AS}. The exclusion process is one of the best studied stochastic interacting particle systems far from equilibrium. It often serves as a reference model for stochastic transport and is exactly solvable \cite{bib:BE}, even in some cases of varying system size \cite{bib:Heilmann,bib:Arita,bib:ALS}.

Beyond the length regulation of a single MT, it is important to investigate the organization of the MT cytoskeleton. A centrosome serves as the microtubule organizing center (MTOC) in animal cells, and a spindle pole body does in yeast cells. The MTOC should be localized at an appropriate position e.g. for a successful cell division. Pushing \cite{bib:ZGRJ,bib:DYC,bib:Gluncic} and pulling \cite{bib:KO,bib:KSN,bib:KK,bib:Laan,bib:TKM} forces, and a combination of both \cite{bib:PLMDJ,bib:MLDPJ} have been considered to lead to an appropriate localization. So far the positioning problem has been addressed by analyzing the mechanical forces applied to the MT network. In this work, we concentrate on the MTOC localization, which is tuned by Kip3p. We introduce a simple stochastic model consisting of the MTOC, two MTs, and Kip3p motors. The system is confined to a one-dimensional space with finite size. Each of the MTs is a segment of the simple exclusion process, which is polymerized and depolymerized. We suppose that polymerizations at the plus ends of the MTs induce stochastic movements (fluctuation) of the MTOC. We show that this simple setting spontaneously causes a bias in the dynamics of the MTOC towards the middle of the cell. We also realize spatially asymmetric positioning by imposing asymmetric parameters as well.

\section{Model}
\begin{figure}
\begin{center}
 \includegraphics[width=0.5\columnwidth]{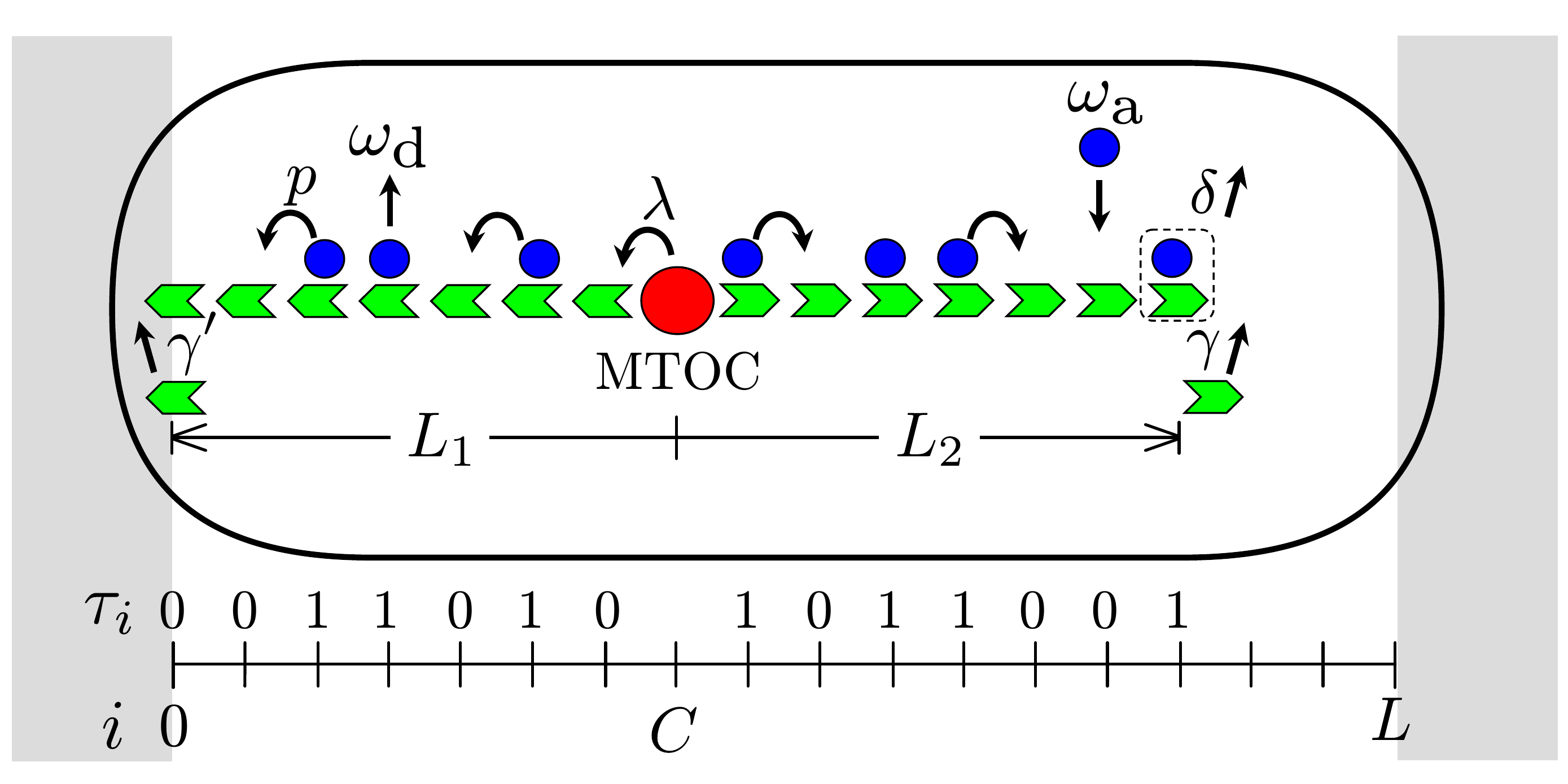}
\caption{Schematic of our model. Each arrow represents a possible stochastic event with the corresponding transition rate. The MTOC position is a discrete variable in the interval $ 0 \le C \le L $. 
\label{fig:schema}}
\end{center}
\end{figure}

We describe a pair of MTs which are connected by a MTOC (see Fig.~\ref{fig:schema}). A notation of discrete, finite one-dimensional coordinates $ \mathbb A = \{ 0,1,\dots, L \} $ is introduced, which captures all available sites inside the cell ($L $ corresponds to the diameter of the cell). For simplicity, we suppose that the MTOC and tubulin dimers are of identical size, and their positions are specified by a number in $ \mathbb A $. We denote the position of the MTOC at time $t$ by $ C (t) \in \mathbb A $. Two MTs ($n=1,2$) are connected to the MTOC, and their lengths $ L_n(t) $ are given by the numbers of tubulin dimers. Therefore the positions of the tips (the plus ends) of the right and left MTs are given as $ C(t)-L_1(t) $ and $ C(t)+L_2(t) $, respectively. Obviously, we have the restriction \begin{align} L_1(t) + L_2(t) \le L . \end{align} The case $ L_1(t) =C(t)$ [resp. $ L_2(t) = L - C(t) $] corresponds to the left (resp. right) tip touching the cell membrane. We also suppose that each tubulin dimer at $i\in \mathbb A$ is either occupied by a kinesin ($ \tau_i=1 $) or empty ($ \tau_i=0 $).

The dynamics of the MT lengths are determined by polymerization and depolymerization at the plus ends. As long as the MT $n=1$ (resp. $n=2 $) does not touch the boundary, i.e. when $ L_1(t) < C(t) $ [resp. $ L_2(t) < L - C(t) $] is satisfied, a tubulin dimer is added with rate $ \gamma $ at the tip of the MT. Polymerization of one tubulin dimer increases the length of the given MT by one unit, i.e. $L_n = \ell \mapsto \ell +1 $. If one of the two tips touches the boundary but the other one does not, polymerization of the former shifts the whole of the system rightward [polymerization of MT $ n=1 $, $ L_1(t) = C (t) $] or leftward [polymerization of MT $ n=2$, $ L_2(t) = L-C (t) $]. For example, the left MT is touching the boundary in Fig.~\ref{fig:schema}; when a tubulin dimer is added to the left MT, the system is shifted rightward, and the configuration changes as $ \tau_0\tau_1\tau_2\cdots =001101\cdots \mapsto 000110\cdots $. We denote the rate for this polymerization with pushing by $ \gamma' $, i.e. the events $ ( L_1 ,C ) = ( \ell, \ell ) \mapsto ( \ell+ 1 , \ell +1) $ and $ ( L_2 ,C ) = ( \ell , L-\ell ) \mapsto ( \ell+1 ,L-\ell -1) $ occur with rate $ \gamma' $. When $ L_1(t) + L_2(t) = L $, i.e. the MTs and MTOC cover all the positions over $\mathbb A$, no polymerization is possible. The depolymerization dynamics is simpler, i.e the tip of each MT is depolymerized $ L_n(t)=\ell \to \ell -1 $ with rate $ \delta $, if it is occupied by a kinesin \cite{bib:MRF}.

Now we define the dynamics of kinesins. If a tubulin dimer next to the MTOC is unoccupied, a kinesin enters it with rate $\lambda$. Kinesins attempt to hop with rate $p$ ($p=1$ without loss of generality) towards the plus ends, if the preceding tubulin dimer is empty. This mutual exclusion is the basic rule of the exclusion process. When a depolymerization occurs, the kinesin at the tip of the depolymerized MT leaves simultaneously. 

Last but not less important elements of the kinesin dynamics are the Langmuir kinetics \cite{bib:PFF,bib:EJS}. Kinesins enter and leave the system in the bulk of the MTs with rates $\omega_\text{a}$ and $\omega_\text{d} $, respectively. In this work we particularly consider the case where these rates are equal, $ \omega_\text{a} = \omega_\text{d} $, so that density profiles can be predicted in simple analytical forms. This choice of parameters does simplify the analysis of the localization mechanism but does not change its nature. We also use the notation $ \Omega = L \omega_\text{d} $, which should be of the order of the hopping rate chosen to be $p=1$, as is often assumed in exclusion processes with the Langmuir kinetics \cite{bib:EJS}.

\begin{figure}[t]
 \begin{center}
\includegraphics[width=0.4\columnwidth]{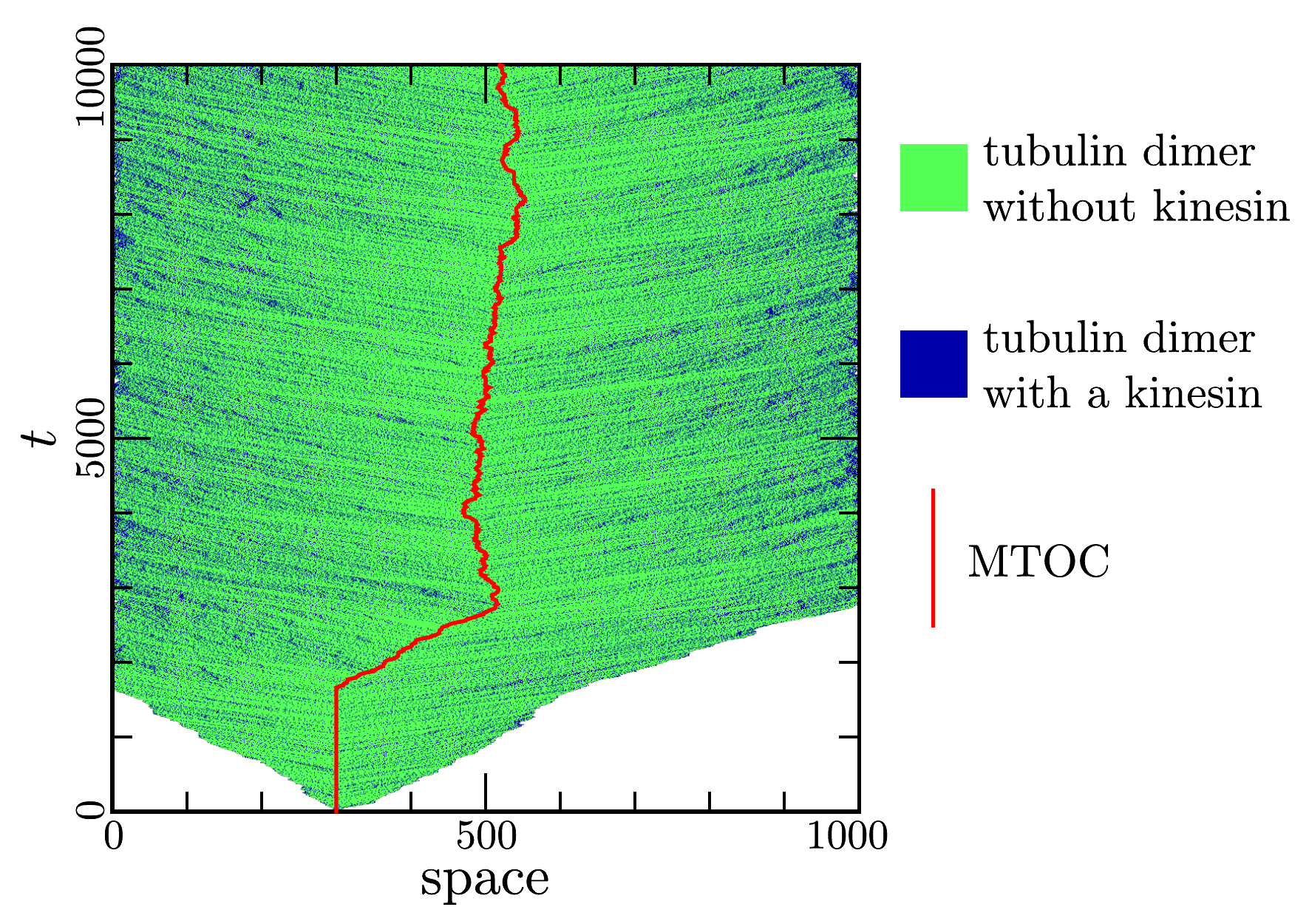}
\caption{Kymograph of the system.
The MTOC is labeled in red, tubulin dimers with (without) kinesin are labeled in blue (green).
We use the initial conditions $ L_1= L_2= 0$ and $ C=300$, and 
the other model-parameters are $ (\lambda,\delta,\gamma,\Omega,L) = (0.1,0.3,0.3,0.4,1000) $.
\label{fig:kymo}}
\end{center}
\end{figure}

\section{Localization mechanism}\label{sec:positioning}
In order to illustrate the localization mechanism we perform simulations with $ \gamma' = \gamma $ for simplicity. 
Figure \ref{fig:kymo} shows a kymograph of the system, starting from the situation where a MTOC is located at some 
point in $ \mathbb A $. As soon as the growing MTs get in contact with the boundaries, a localization starts. We observe that the MTOC rapidly moves to the middle of the system, where it stays localized. 

From the left-right symmetry of our model, the average value of the rescaled MTOC position $ c=C/L $ has to be 
$\langle c\rangle = \frac 1 2 $. The actual degree of MTOC localization is rather quantified by the standard deviation 
of the spatial distribution 
\begin{align}
 \sigma = \sqrt{ \langle c^2 \rangle - \langle c \rangle^2 }\ , 
 \end{align}
which serves as an order parameter for the localization transition. We say that the MTOC is localized, when it is centered in the following stronger sense: 
\begin{align} \label{eq:Deltac=0}
 \lim_{ L\to\infty } \sigma =0 .
\end{align}

\begin{figure}[t]
\begin{center}
 \includegraphics[width=38mm]{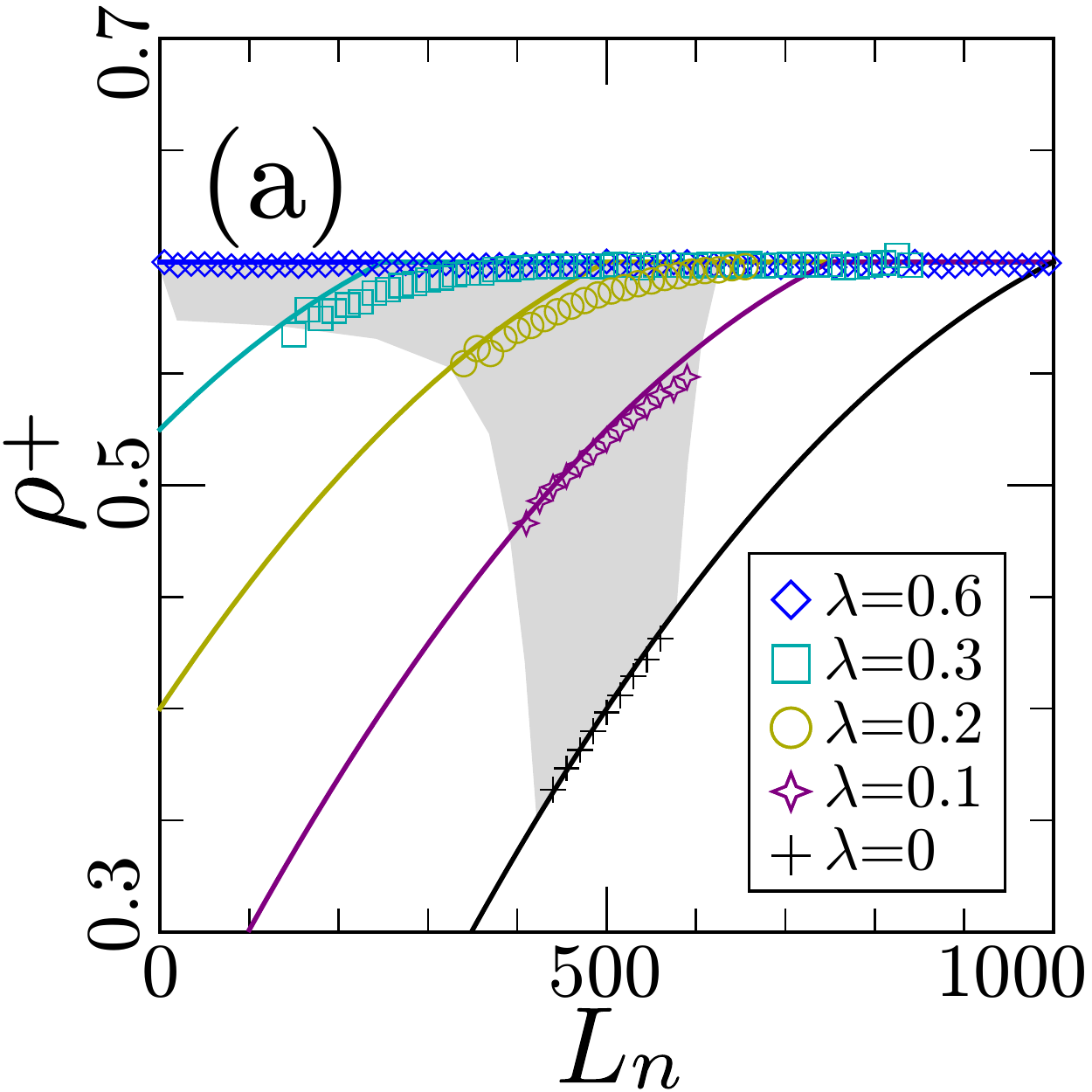}\quad 
 \includegraphics[width=38mm]{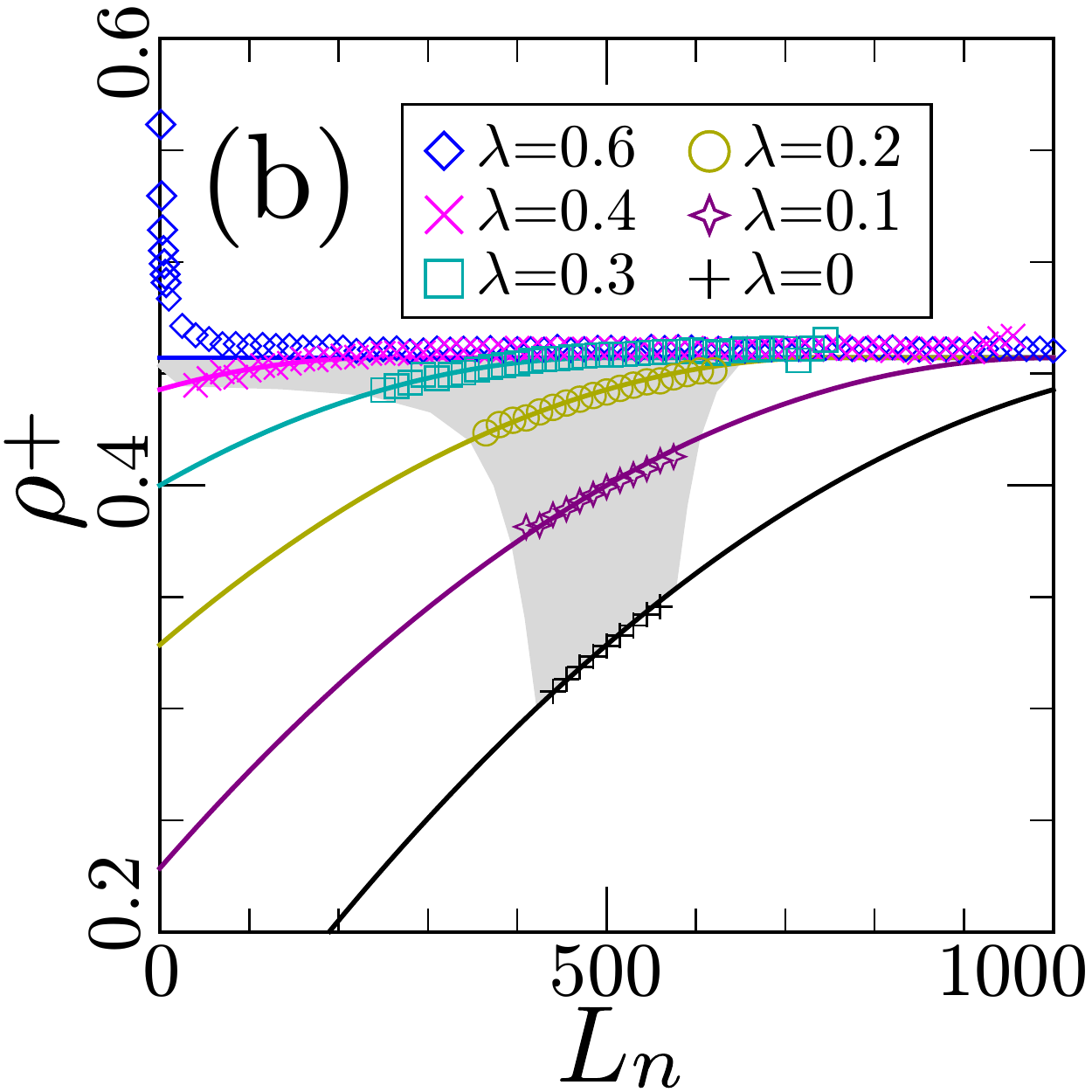}
\caption{Tip densities depending on MT lengths $ L_n $ for $ \delta = 0.4 $ (a) and $ \delta = 0.7 $ (b). The values of $ \lambda $ are indicated in the panels, and the other parameters were set as $ ( \gamma=\gamma',\Omega ,L) = ( 0.3,0.4,1000 ) $. The plot markers are simulation results, and the lines correspond to the prediction \eqref{eq:relation}. 
Due to the localization of the MTOC, reliable simulation results outside the shaded area are difficult to obtain. 
\label{fig:tip-density}}
\end{center}
\end{figure}

In the following, we explore the mechanism underlying the MTOC localization. 
Let us denote by $ \rho^+_{L_n} $ the tip (i.e. $+$ end) densities of the two MTs $n=1,2$.
We shall see that the dependency of these quantities on the lengths $ L_n $
 plays an important role in explaining the localization mechanism.
When $ \gamma $ is large enough, the $+$ ends of both two MTs often touch the boundaries, 
i.e. $ L_1 \approx C, L_2 \approx L-C $. The MTOC is more likely pushed rightward, if 
$ \rho^+_{ L_1 } < \rho^+_{ L_2 } $ (and vice versa), since the depolymerization 
 rate $ \delta \rho^+_{L_2} $ of the right MT is higher than the left one $ \delta \rho^+_{L_1} $. 
Therefore the localization holds when the following condition is satisfied near $ C= L/2 $: 
\begin{align}\label{eq:condition}
 L_1 < L_2 \Rightarrow \rho^+_{ L_1 } < \rho^+_{ L_2 } , \ 
 L_1 > L_2 \Rightarrow \rho^+_{ L_1 } > \rho^+_{ L_2 } . \ 
\end{align}

In order to calculate the tip densities, we consider macroscopic density profiles over the two MTs, i.e. 
\begin{align}\label{eq:rho(x)}
 \rho_1 (x) = \langle \tau_{x L } \rangle\ ( x < c ), \ 
 \rho_2 (x) = \langle \tau_{x L } \rangle\ ( c < x ) .
\end{align}
In the limit $ L\to \infty $ and for $ \omega_\text{a} = \omega_\text{d} $,
the density profiles are described by the hydrodynamic equations \cite{bib:PFF,bib:EJS} 
\begin{align}\label{eq:rho(x)}
 (1-2\rho_1) ( \partial_x \rho_1 + \Omega ) = 0 , \ 
 (1-2\rho_2) ( \partial_x \rho_2 - \Omega ) = 0 ,
\end{align} 
with the rescaled position $ x = i/L $. The solutions of these equations become linear or piecewise linear 
\cite{bib:EJS} , depending on the parameters and the MTOC position.
Here we focus on the low density (LD) profiles
\begin{align}
\label{eq:LD}
 \rho^\text{LD}_1(x) \simeq \lambda - ( x - c ) \Omega , \ 
 \rho^\text{LD}_2(x) \simeq \lambda + ( x - c ) \Omega , 
\end{align} 
 which is the most relevant for MTOC positioning. 

The tip densities are, in general, different from the limits of the macroscopic densities 
$ R_{ L_1 }:= \rho_1 ( 0) $ and $ R_{ L_2 }:= \rho_2 ( 1 ) $, since boundary layers can exist (see the Appendix). However, 
 the tip densities can be obtained from macroscopic considerations: The particle current at the tip is given by 
 $\rho^+_{L_n} \delta$ and should approximately agree to the bulk current close to the boundary $R_{ L_n } ( 1 - R_{ L_n } ) $, hence the relations 
\begin{align}\label{eq:relation}
 \rho^+_{L_n} \approx R_{ L_n } ( 1 - R_{ L_n } ) / \delta\quad 
 (n=1,2). 
\end{align}
One finds that the condition \eqref{eq:condition} is satisfied only in the case where the low densities \eqref{eq:LD} cover all over $ \mathbb A $, see the Appendix. In this case, the expected approximation formula becomes 
\begin{align} \label{eq:rho+Ln}
 \rho^+_{L_n} \approx 
 ( \lambda + \Omega L_n / L ) ( 1 - \lambda - \Omega L_n / L ) / \delta, 
\end{align}
by substituting $ \rho^+_{L_1} = \rho^\text{LD}_1(0) $ and $ \rho^+_{L_2 } = \rho^\text{LD}_2(1) $
into Eq.~\eqref{eq:relation}.

 It is remarkable that $ \rho^+_{L_n} $'s are controlled by the injection rate $ \lambda $ in this case. 
We actually observe nonzero gradients in the vicinity of $ L_n = L/2 $ in Fig. \ref{fig:tip-density}
[in the cases of $ \lambda \le 0.2 $ for (a) $ \delta =0.4 $, and $ \lambda \le 0.3 $ for (b) $ \delta =0.7 $], 
realizing the condition \eqref{eq:condition}. 

\begin{figure}
\begin{center}
 \includegraphics[width=45mm]{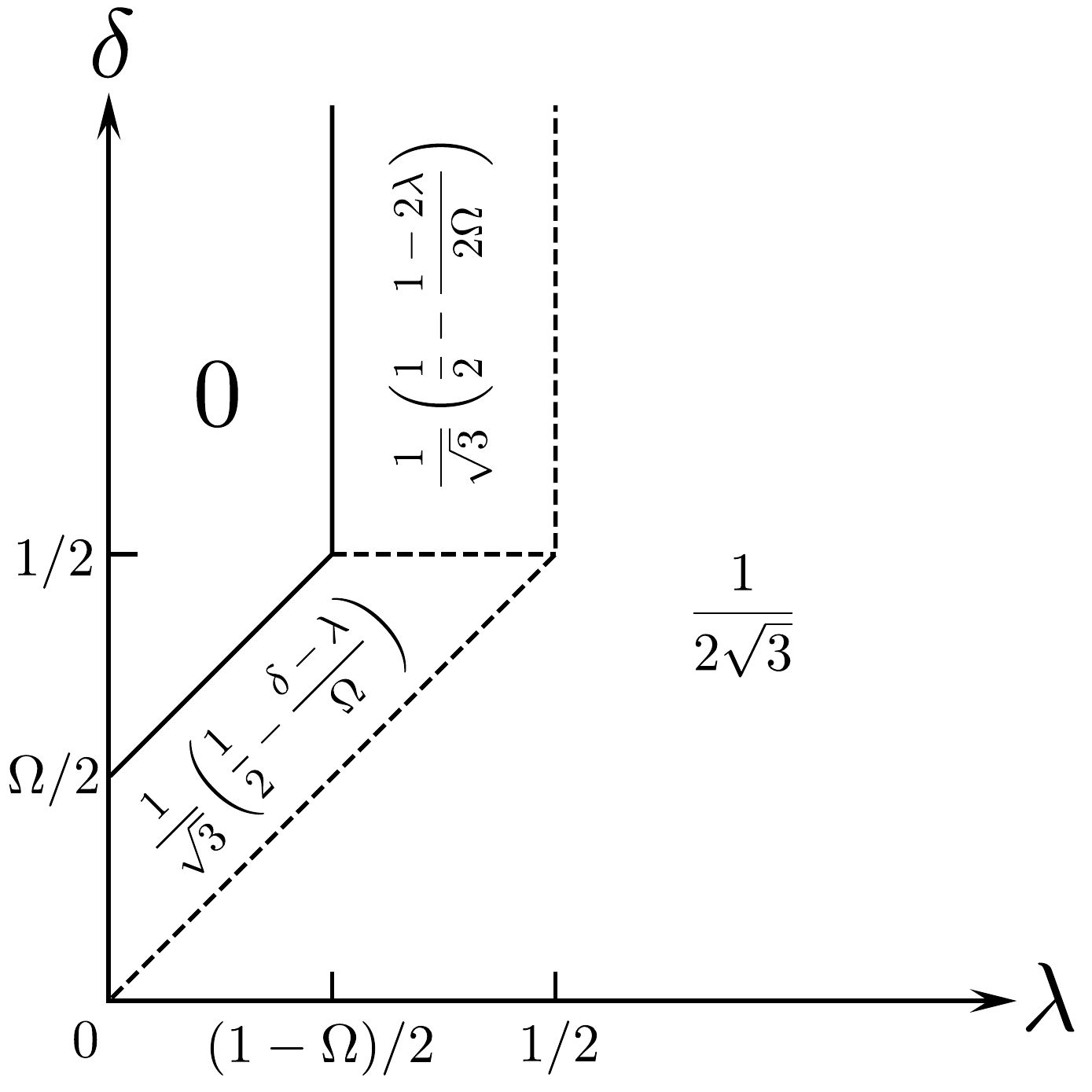}
\caption{Phase diagram of $ \sigma $ characterizing the localization of the MTOC. 
The Langmuir rate is $ 0 < \Omega < 1 $, and the polymerization rate $ \gamma $ is larger than $ \frac 1 4 $. In each region, the predicted value of $ \sigma $ in the limit $ L\to\infty $ is indicated. In particular ``0'' corresponds to the localization phase. 
\label{fig:phase-diagram}}
\end{center}
\end{figure}

We identify the parameter region, where one observes a low density profile \eqref{eq:LD} on both MTs if the MTOC is positioned in the vicinity of $c = \frac 1 2 $ as 
\begin{align}\label{eq:localization-phase}
 \lambda < ( 1-\Omega )/2 \ \wedge \ 0 < \Omega < 1 \ \wedge \ \delta > \Omega / 2 +\lambda . 
\end{align}
We refer to this regime as \textit{localization phase} (denoted by 0 in Fig. \ref{fig:phase-diagram}). Violating these conditions leads to a macroscopic domain of density $1/2$ or higher near the MT plus ends. For a set of parameters in the \textit{delocalization phase} (i.e. outside the localization phase), the tip densities are constant in the vicinity of $ L_n = L/2 $, see Fig. \ref{fig:tip-density} [$ \lambda > 0.2 $ for (a) $ \delta =0.4 $, and $ \lambda > 0.3 $ for (b) $ \delta =0.7 $]. 

\section{Dynamics of the MTOC position}
\begin{figure}
\begin{center}
 \includegraphics[width=38mm]{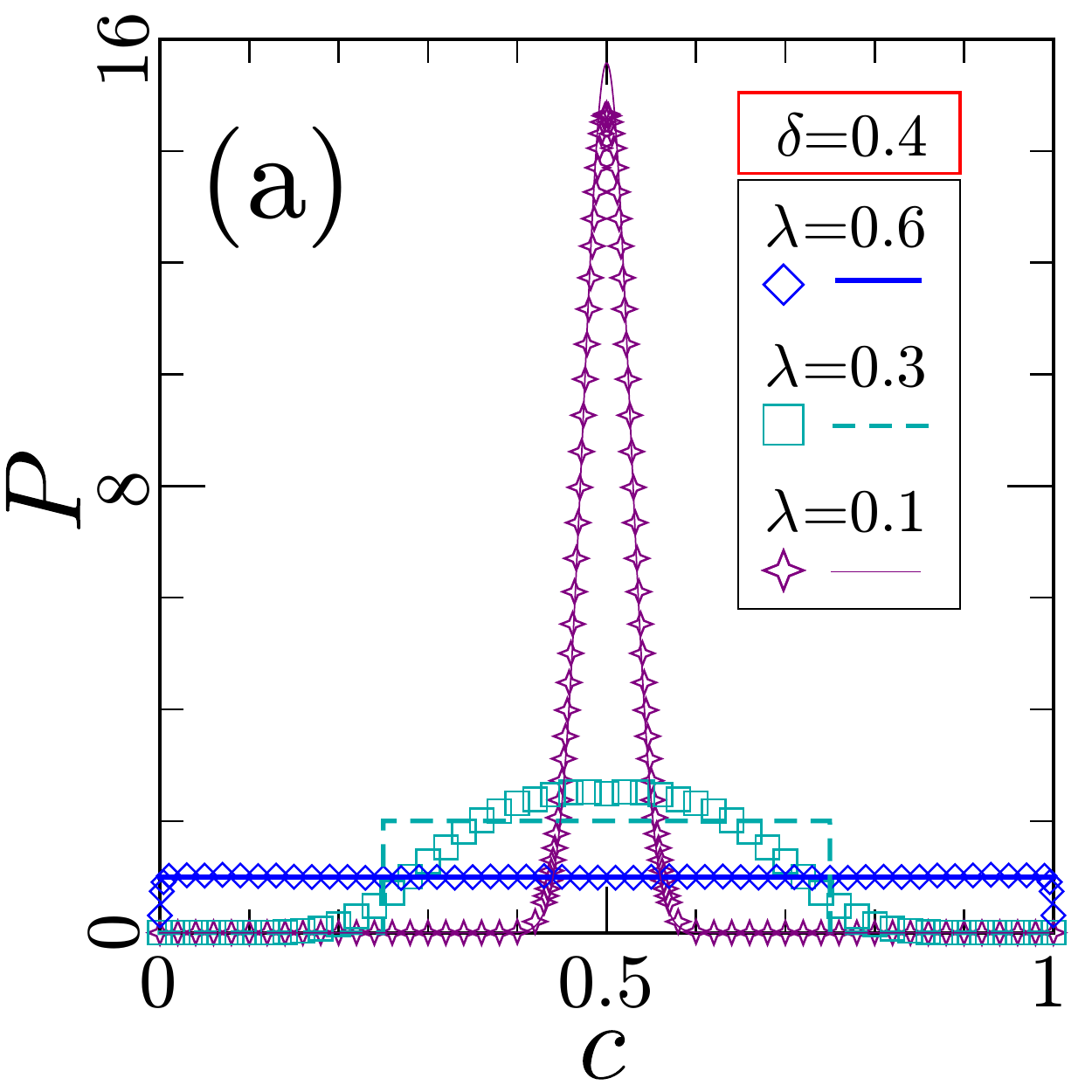}
 \includegraphics[width=38mm]{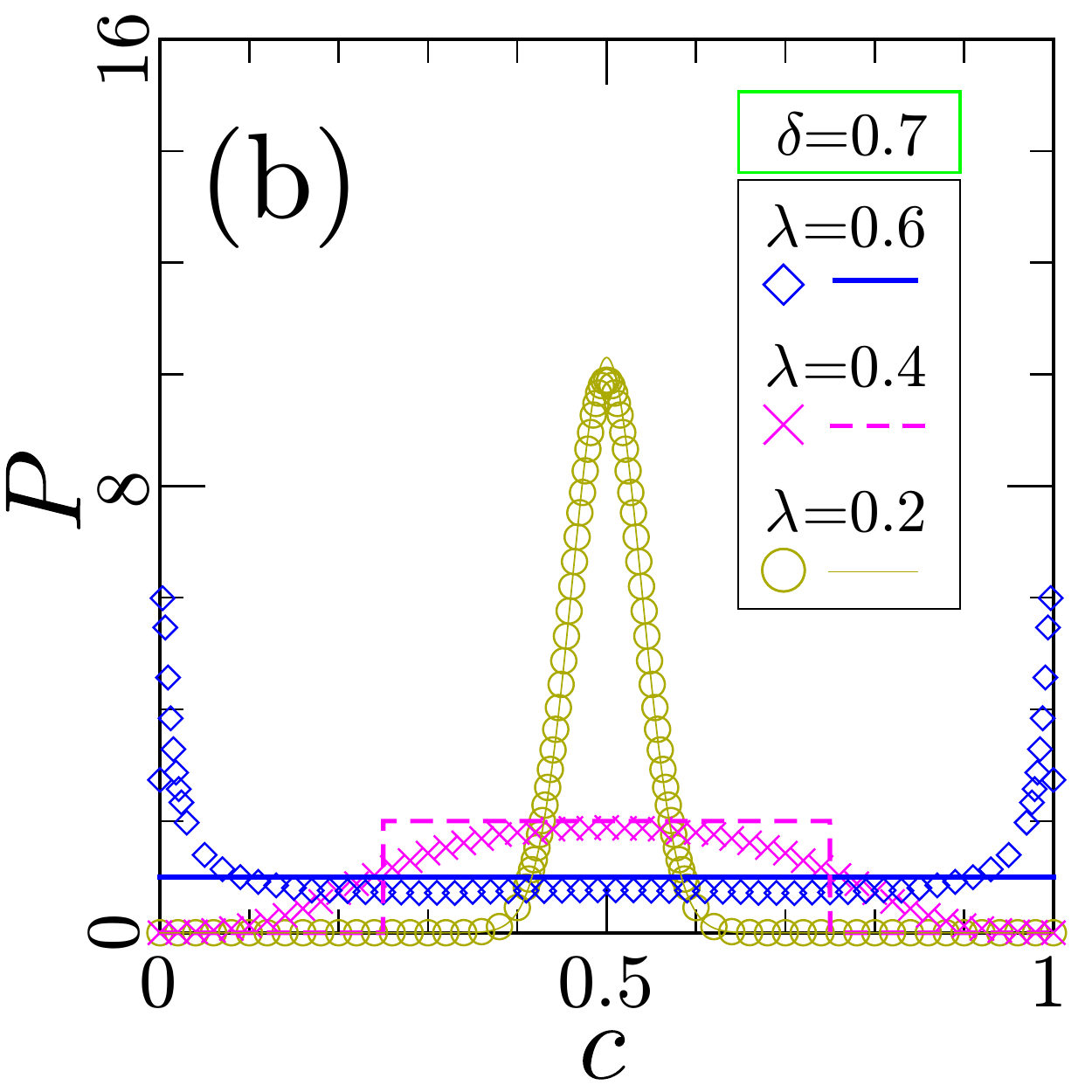}
 \includegraphics[width=38mm]{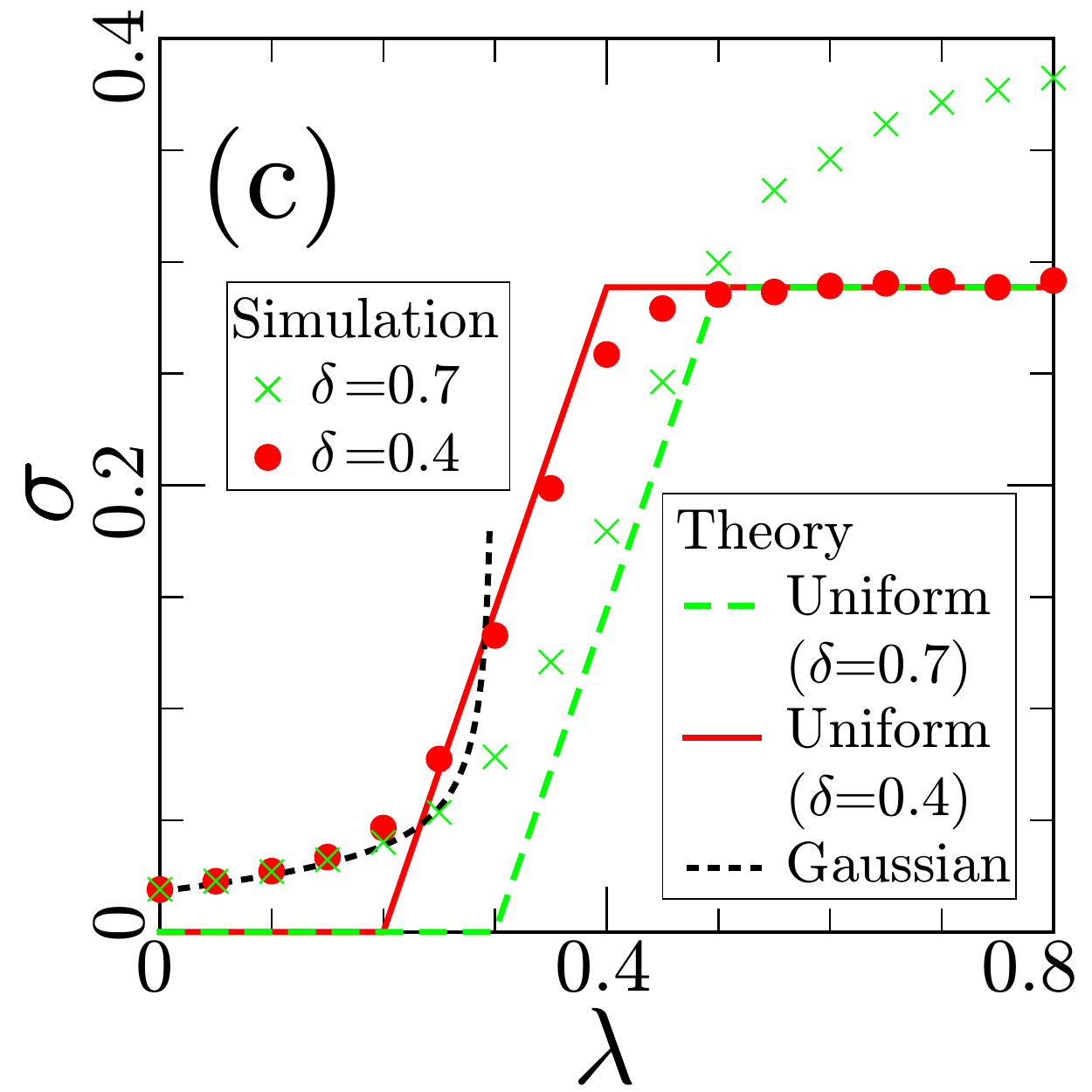}
 \includegraphics[width=38mm]{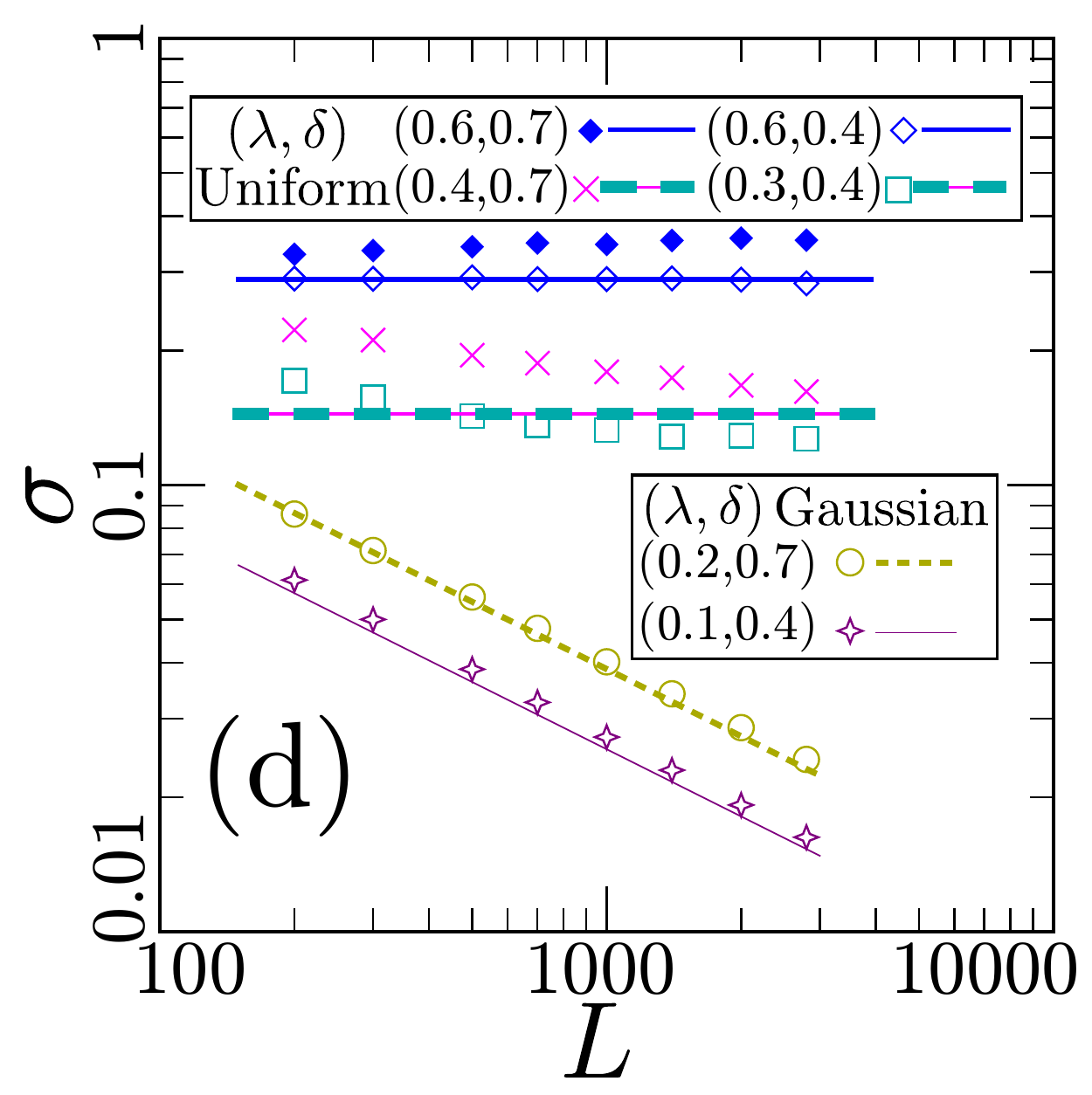}
\caption{Probability distribution $ P(c) $ of the MTOC position (a,b)
and its standard deviation $ \sigma $ (c,d).
The Langmuir and polymerization rates are 
 $ ( \Omega ,\gamma ) = ( 0.4 , 0.3) $, and the input and depolymerization rates $ (\lambda,\delta) $
 are indicated in the panels.
For the system size for (a,b,c), we have set $ L=1000 $. 
The theoretical lines in (a,b) are 
the Gaussian distribution \eqref{eq:P(c)=} with \eqref{eq:1/sigma^2=}, 
or the uniform distribution $ P(c) = \frac{1}{u-v} $ with \eqref{eq:u=}. 
The theoretical lines in (c,d) correspond to 
Eq.~\eqref{eq:1/sigma^2=} for the Gaussian,
or $\frac{u-v}{2\sqrt{3} }$ with Eq.~\eqref{eq:u=} for the uniform. 
 \label{fig:P-sigma}}
\end{center}
\end{figure}

We now discuss a phenomenological approach to the dynamics of $ C $. We consider that the motion of the MTOC is effectively governed by a random walk with leftward and rightward hopping rates, $ q \rho^+_{ C } $ and $ q \rho^+_{ L - C } $, respectively, with some factor $q$. This interpretation leads to the stationary distribution as 
\begin{align}
 P(C) = 
 \frac{1}{Z} \prod_{i=0}^{ | C-L / 2 | } 
 \frac{ \rho^+_{ L / 2 + i } }{ \rho^+_{ L / 2 - i -1 } } 
\end{align}
with normalization $Z$. Furthermore this expression is approximated in the localization phase as 
\begin{align}
\label{eq:P(c)=}
& P( c ) \simeq 
 \frac{ 1 }{ \sqrt{ 2 \pi \sigma ^2 } } \exp \Big[ - \frac{1}{2\sigma^2 } \Big( c - \frac 1 2 \Big) ^2 \Big] , \\ 
\label{eq:1/sigma^2=}
& \frac{1}{ \sigma^2} \approx 
 2 L \frac{ \partial }{\partial x } \ln \rho^+_{ x L } \Big|_{ x=\frac 1 2 }
 = \frac{ 8 \Omega ( 1- 2 \lambda - \Omega ) L }{ (2 \lambda + \Omega ) (2 -2 \lambda - \Omega ) } 
\end{align}
 in the limit $ L \to \infty $, due to Eq.~\eqref{eq:rho+Ln}. We find good agreements between simulations and this Gaussian form in Fig.~\ref{fig:P-sigma} (a,b), and we observe $ \sigma = O( 1 / \sqrt{ L } ) $ in Fig.~\ref{fig:P-sigma} (c,d).

On the other hand, the MTOC position is \textit{confined} to a finite interval $ [ u , v ] $ in the delocalization phase, as long as $ \rho^+_{L_1} = \rho^+_{L_2} $ is satisfied. One finds the values of $u$ and $ v=1-u $ as 
\begin{align}\label{eq:u=}
 u = \begin{cases}
 \frac{ \delta - \lambda }{ \Omega } 
 & \text{($ \delta \le \frac 1 2 \wedge \delta + \frac \Omega 2 < \lambda < \frac{1}{2} $)}, \\
 \frac{ 1 - 2\lambda }{ 2\Omega } 
 & \text{($ \delta > \frac 1 2 \wedge \frac{1-\Omega}{2} < \lambda < \frac{1}{2} $)} ,\\
 0 & \text{(otherwise)} , \\
 \end{cases}
\end{align}
see the Appendix. When we assume that the MTOC is uniformly distributed over $ [u,v] $, the standard deviation is $ \frac{ u-v}{ 2\sqrt 3 } $ in the limit $L\to\infty $. The results of the three subphases are summarized in Fig. \ref{fig:phase-diagram}. We compare the prediction with simulations, as shown in Fig.~\ref{fig:P-sigma}. We observe that the MTOC is often trapped at the boundaries for large $ \lambda $ and $ \delta $ [Fig. \ref{fig:P-sigma} (b)], which cause a quantitative disagreement of $ \sigma $, although qualitatively correct behaviors $ \sigma =O(L^0) $ are found [Fig.~\ref{fig:P-sigma} (c,d)].

\begin{figure}
\begin{center}
 \includegraphics[width=38mm]{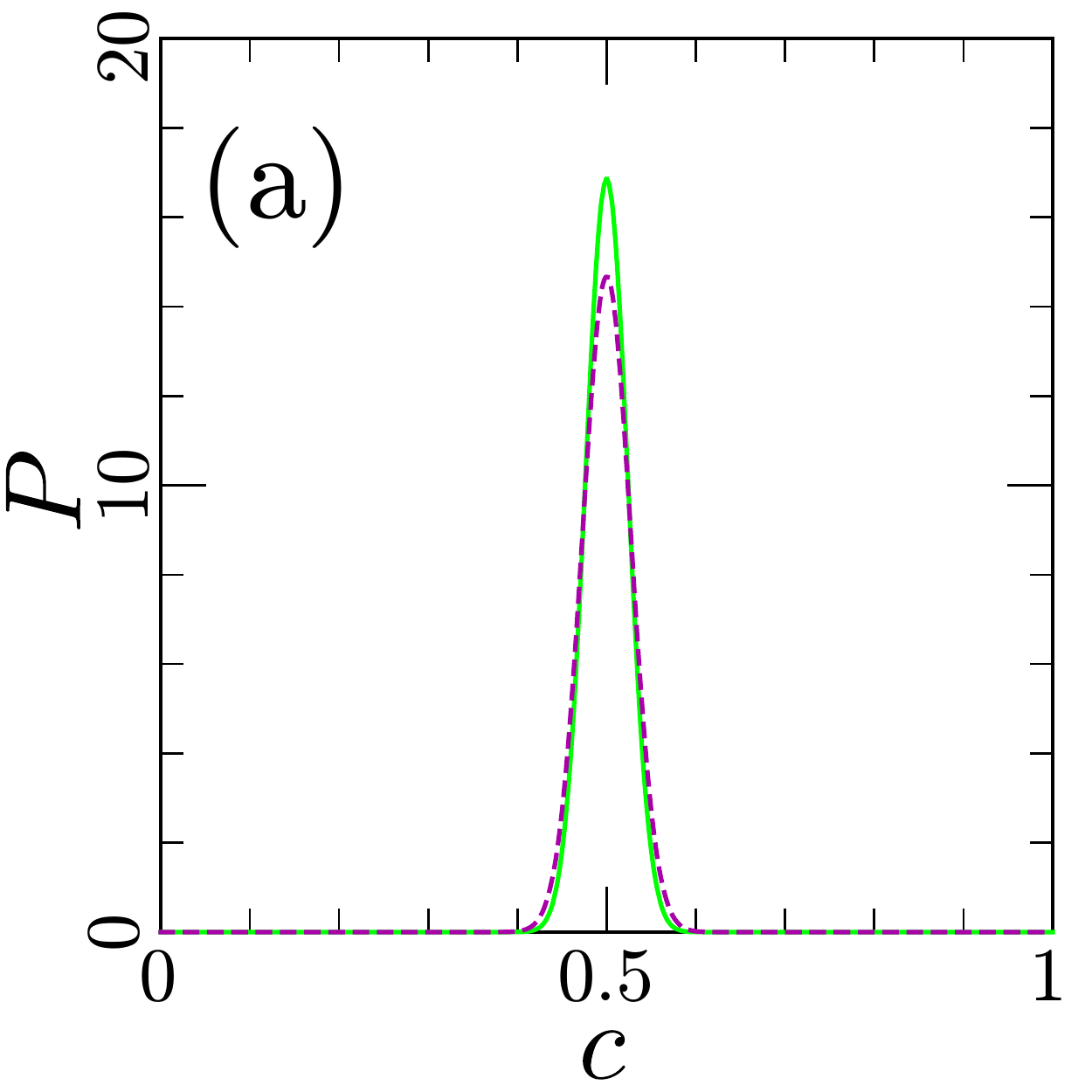} 
 \includegraphics[width=38mm]{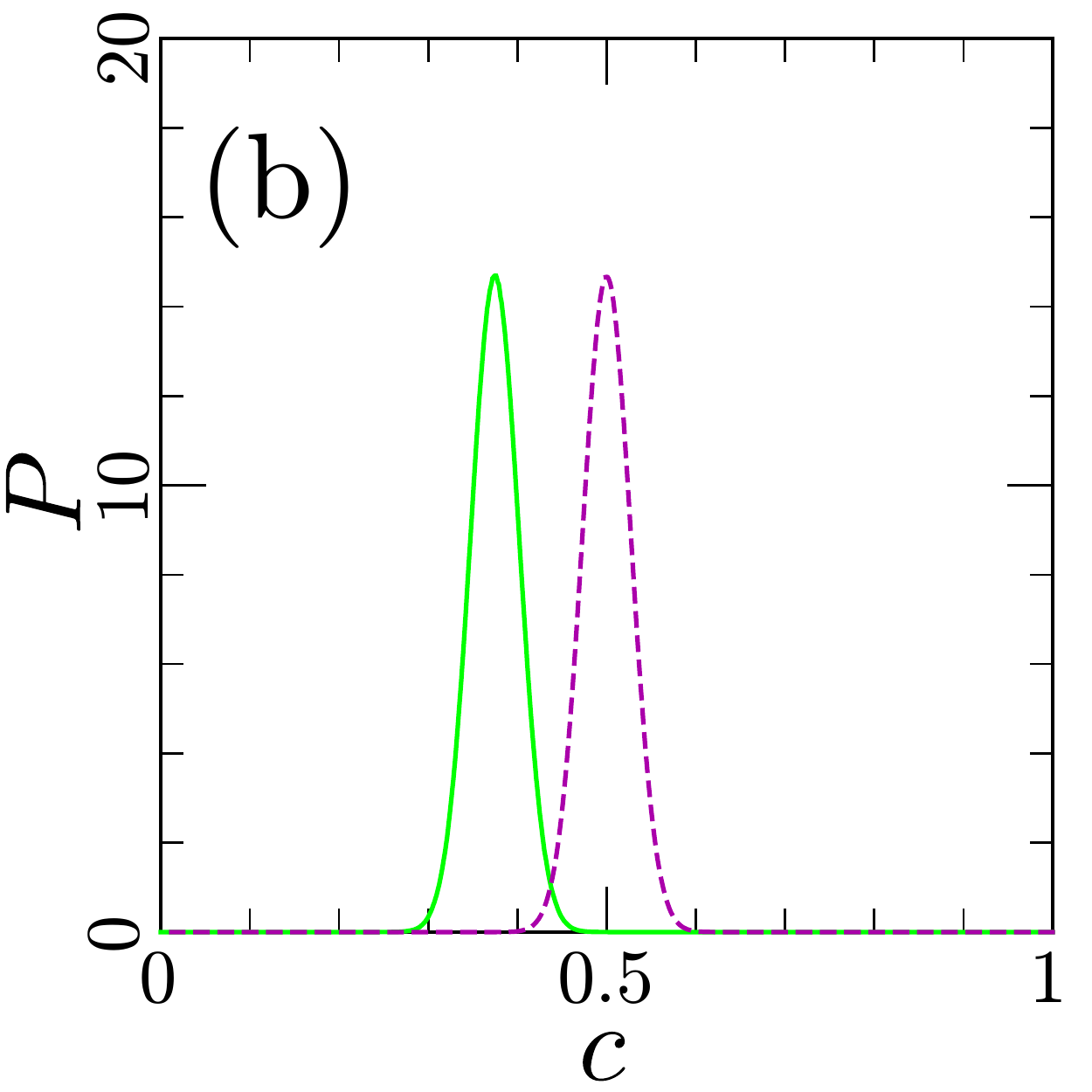} 
 \includegraphics[width=38mm]{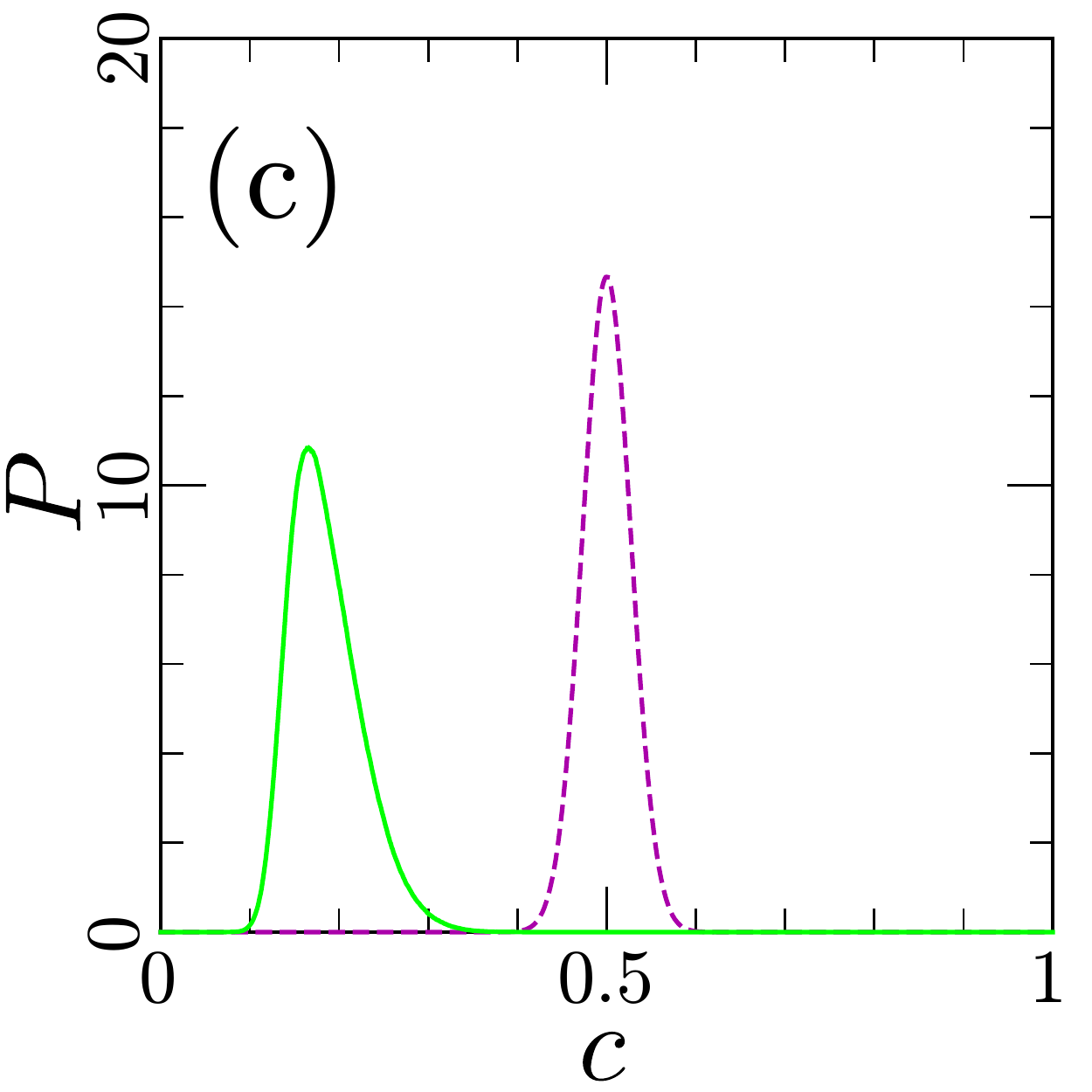} 
 \includegraphics[width=38mm]{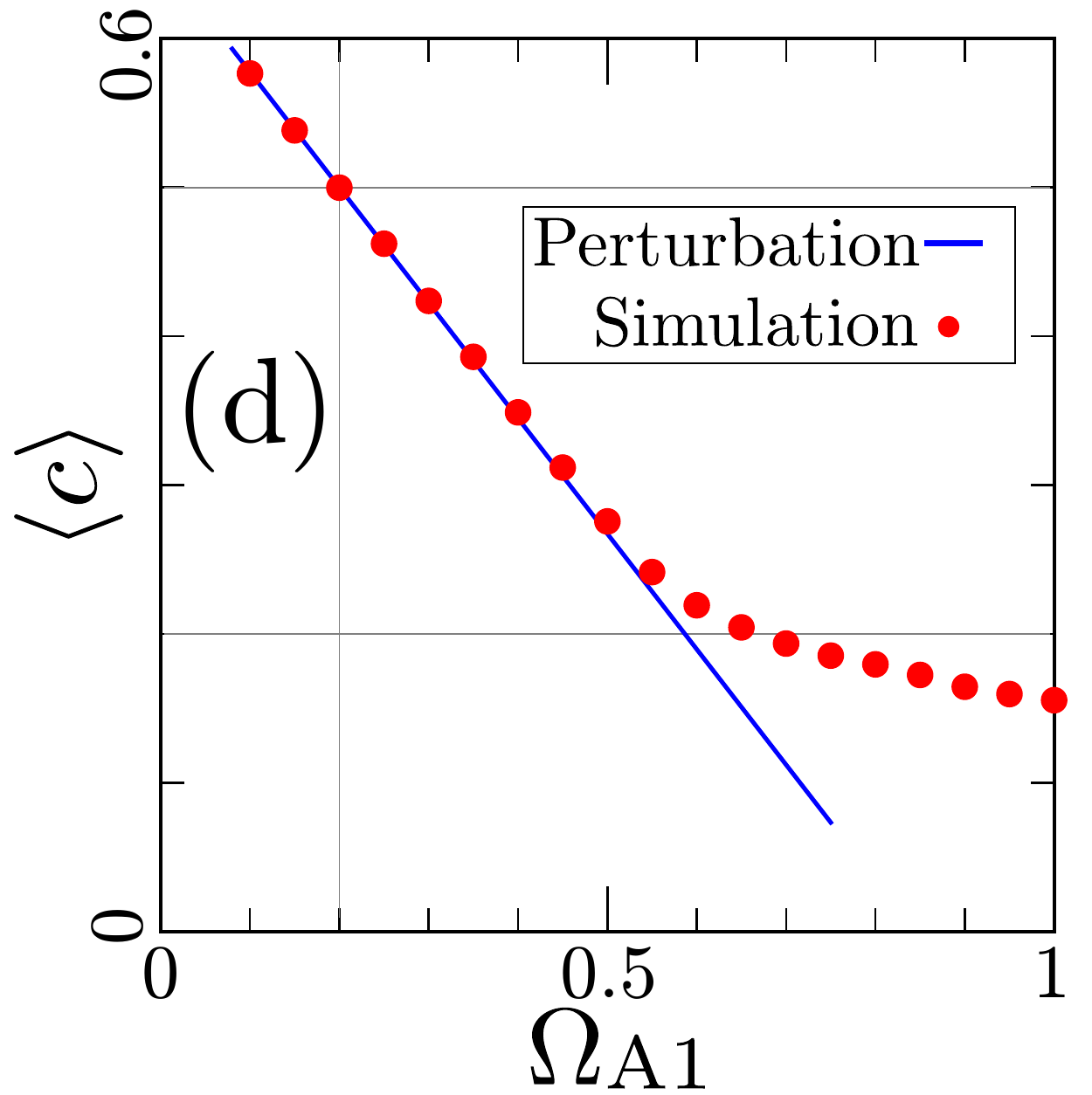} 
\caption{(a,b,c) Distributions $ P(c) $ of the MTOC position in generalized models. 
We drew the dashed line of the original problem 
with $ ( \lambda, \delta,\gamma=\gamma' , \Omega_\text{A} = \Omega_\text{D} ,L ) =
 (0.1 , 0.4 , 0.3 , 0.4, 10^3) $ in each panel. 
From this reference case, we changed the parameters as 
 (a) $ \gamma' = 0.5 $,
 (b) $ (\lambda_1, \lambda_2) = ( 0.15, 0.05 ) $, and 
 (c) $ ( \Omega_\text{A1}, b )= ( 1.5 ,0.2) $,
 corresponding to solid lines. 
 (d) Average MTOC position 
 in the model with varying $ \Omega_\text{A1} $
 for $ b=0.2 $. We also plot the first order perturbation, Eq.~\eqref{eq:first-order} for comparison. 
\label{fig:robustness}}
\end{center}
\end{figure}

\section{Generalizations}
We now generalize our model and consider robustness; e.g. it is natural to consider the case $ \gamma'<\gamma $.
The localization in the middle of the cell is indeed realized with a sharper distribution, see Fig.~\ref{fig:robustness} (a). 

It is important to discuss spatially asymmetric positioning, as observed in budding yeast. In our model, we impose different input rates of motors at different MTs $ \lambda_1 ,\lambda_2 $. Figure~\ref{fig:robustness} (b) shows that the asymmetry of input rates $ \lambda_1 > \lambda_2 $ shift the MTOC position leftward. 

One can also assign different attachment rates to two compartments in a cell. Kinesin motors attach to the MTs with rates $ \Omega_\text{A1} $ and $ \Omega_\text{A2} $ for sites $ i < b L $ and $i\ge b L $, respectively, with some $0< b<1 $. This arrangement models different kinesin concentrations. For simplicity the attachment rate $ \Omega_\text{A1} $ is different from $ \Omega_\text{D} $, but we keep $ \Omega_\text{A2 } = \Omega_\text{D} =: \Omega $. 
One finds an asymmetric localization by tuning $ \Omega_\text{A1} $ in Fig.~\ref{fig:robustness} (c).
It is possible to perform perturbative calculation near 
 $ \Omega_\text{A1} / \Omega - 1 := \epsilon =0 $,
and it turns out that 
\begin{align}\label{eq:first-order}
 \langle c\rangle \simeq \frac{1}{2} - \frac{ \epsilon}{ 8\Omega } 
 \bigg[ 2 b\, \Omega - \ln \frac{ 2\lambda -1 + \Omega }{ 2\lambda -1+ (1-2b)\Omega } \bigg], 
\end{align}
see the Appendix for details. We observe a good agreement between this formula and simulations in Fig.~\ref{fig:robustness}~(d).

\section{Conclusions}
By using a simple stochastic model for MT growth, we identified a simple, robust mechanism for MTOC positioning by processive Kip3p motors, which enhance MT depolymerization at the plus end. We observe a stable MTOC localization in a parameter regime where the motor densities of the MT tips increase with the lengths of the MTs. In this localization phase, the tip densities are controlled by the input rate of the kinesin motors in proximity of the MTOC. 

We imposed a polymerization rate with the global shift $ \gamma' $, which can be regarded as global fluctuation of the system. The other stochastic events indeed locally occur. We also emphasize that, in our model, one does not need to bias the parameters, depending on the MTOC position at the moment. 

In accordance to experimental observations in yeast the Kip3p motors mainly influence the MT depolymerization close to the membrane, since the MTOC localization takes place in a parameter regime where unconfined MTs steadily grow. This fact implies that MTOC positioning and the length regulation of free filaments depend on different mechanisms.

A quantitative model of motor-induced MTOC positioning should take into account several additional features; the higher-dimensional structure of the MT cytoskeleton, the presence of several MT protofilaments, and the influence of other MT associated protein on the MT dynamics. There is also strong evidence that pushing and pulling forces have some impact on the positioning of the MT network \cite{bib:Gluncic,bib:MLDPJ}. Our results show, however, that positioning can already be achieved by motor-induced deplymerizations.

\subsection*{Acknowledgements}
This work was supported by the SFB 1027. CA and LS thank the University of Warwick for hospitality in the symposium ``Fluctuation-driven phenomena in biological systems''.

\section*{Appendix}

 We first show various types of macroscopic density profiles in the symmetric model, and we also derive the phase boundaries. For the second part, we discuss hydrodynamic equations for the generalized model with different attachment rates. 

We consider the macroscopic density profiles of the two MTs $n=1,2 $: 
\begin{align}\label{eq:rho(x)}
 \rho_1 (x) = \langle \tau_{x L } \rangle\ ( x < c ), \ 
 \rho_2 (x) = \langle \tau_{x L } \rangle\ ( c < x ), 
\end{align}
assuming that $ L_1 $ and $L_2$ are large enough. 
We also assume that each MT is effectively regarded as the usual TASEP with the Langmuir kinetics. 
In the stationary state, the hydrodynamic equations that they obey are given as \cite{bib:PFF,bib:EJS} 
\begin{align}\label{eq:rho(x)}
 (1-2\rho_1) ( \partial_x \rho_1 + \Omega ) = 0 , \ 
 (1-2\rho_2) ( \partial_x \rho_2 - \Omega ) = 0 . 
\end{align} 
The solution to these equations becomes linear or piecewise linear,
i.e. the following three types or combinations of them: the low densities (LD), Eq.~\eqref{eq:LD}; 
the high densities (HD)
\begin{align} 
\label{eq:HD}
 \rho^\text{HD}_1(x) \simeq 1-\delta - x\, \Omega , \ 
 \rho^\text{HD}_2(x) \simeq 1-\delta - (1-x) \Omega ; 
\end{align} 
and the maximal-current densities (MC)
\begin{align}
\label{eq:MC}
 \rho^\text{MC}_1 = 1/2 , \ 
 \rho^\text{MC}_2 = 1/2 , 
\end{align} 
which coincide with the Langmuir density $\frac{ \Omega_\text{A} }{ \Omega_\text{A} + \Omega_\text{D} }$. In particular a discontinuity or a shock (say at $x=s$) appears between a LD and a HD, satisfying 
$ \displaystyle \lim_{x\nearrow s} \rho_n (x) = 1- \displaystyle \lim_{x\searrow s} \rho_n (x) $. 
An important remark in our case is that the shapes of the profiles change depending on the MTOC position as well as the other parameters. Although the full classification of the density profiles in the $\lambda$-$\delta$ space could be done, according to the existing works \cite{bib:PFF,bib:EJS}, here we only consider the phase boundaries drawn in Fig.~\ref{fig:tip-density}.

\begin{figure}
\begin{center}
 \includegraphics[width=0.81\columnwidth]{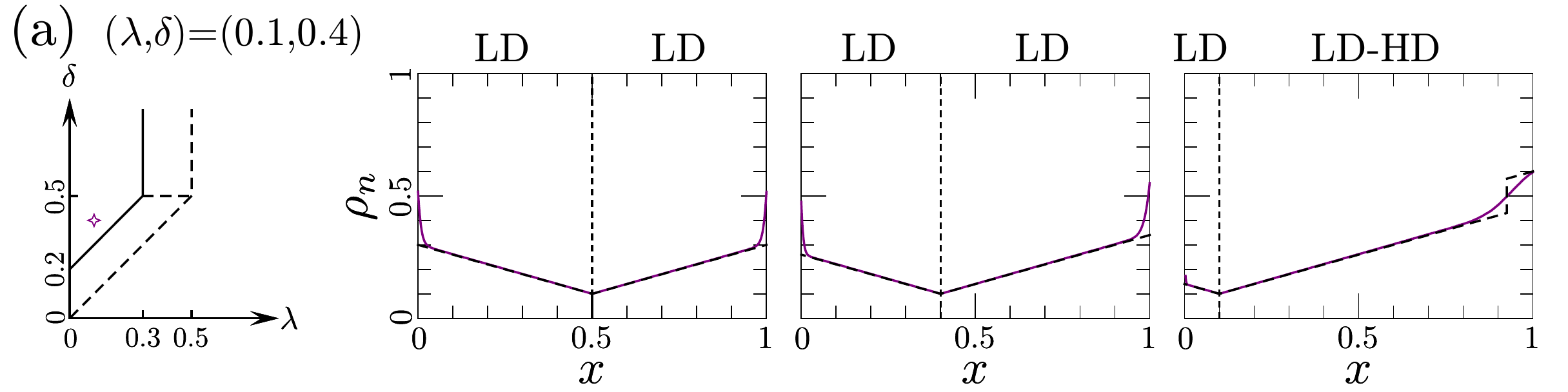}
 \includegraphics[width=0.81\columnwidth]{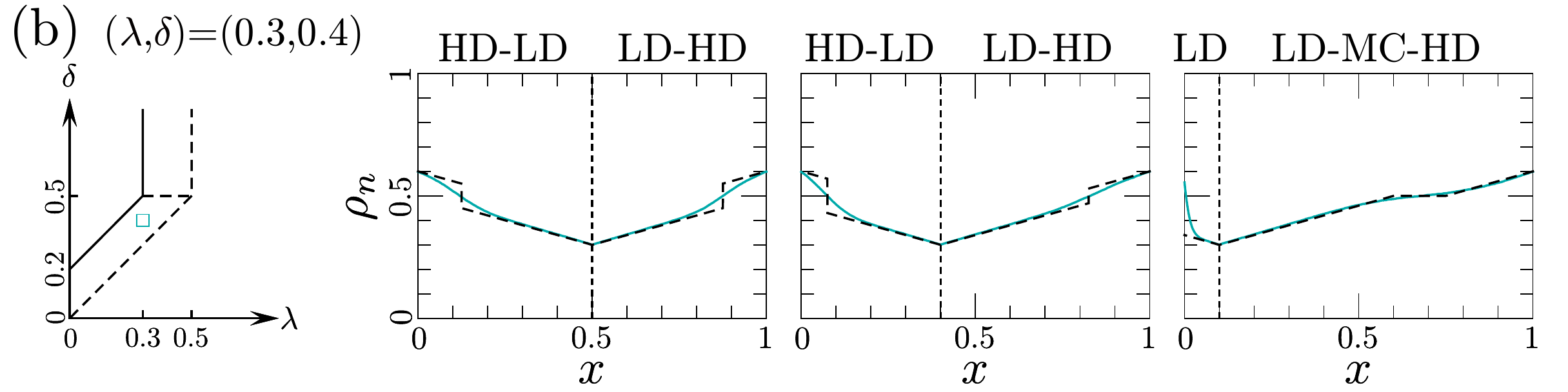}
 \includegraphics[width=0.81\columnwidth]{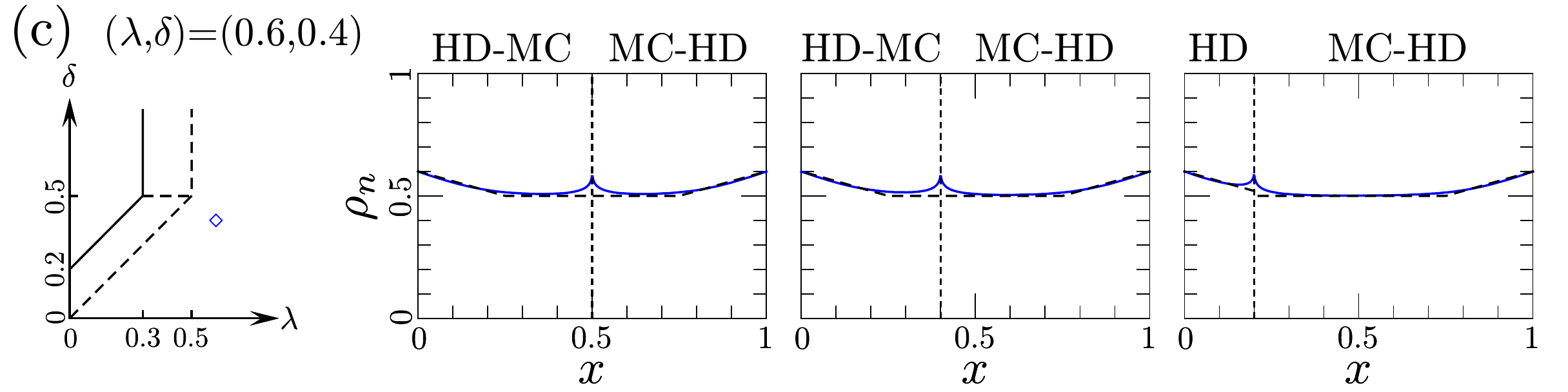}
 \includegraphics[width=0.81\columnwidth]{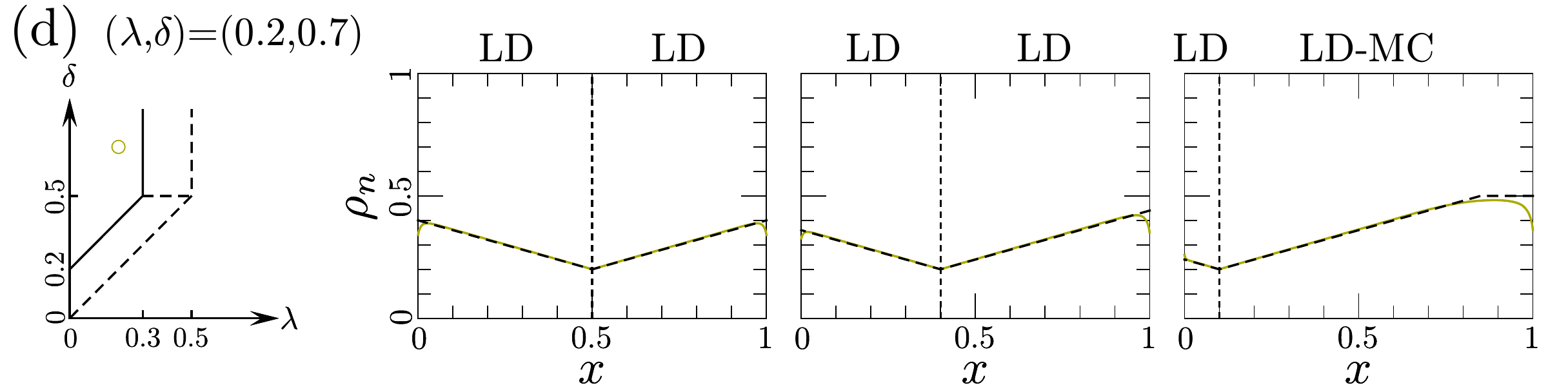}
 \includegraphics[width=0.81\columnwidth]{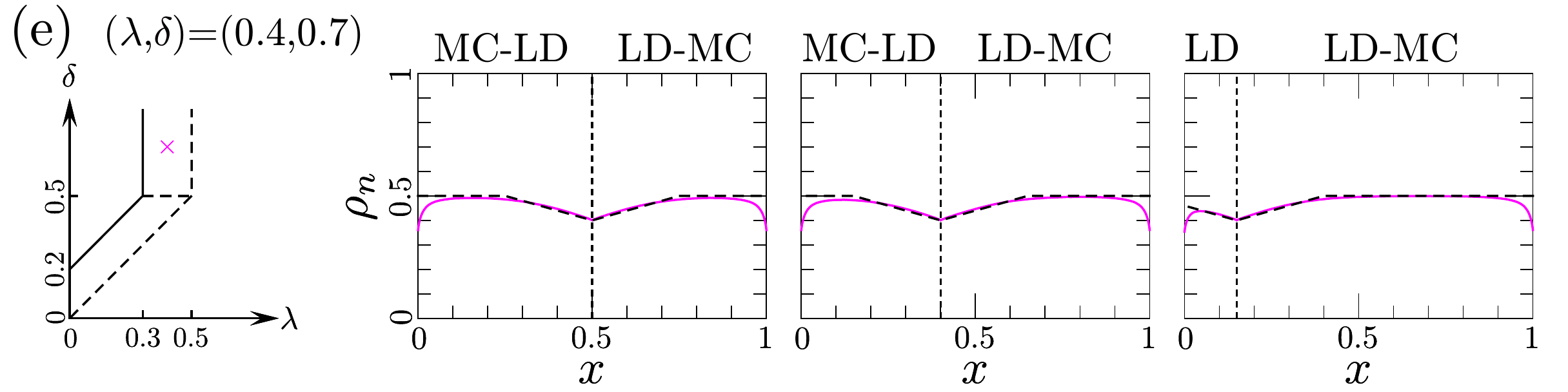}
 \includegraphics[width=0.81\columnwidth]{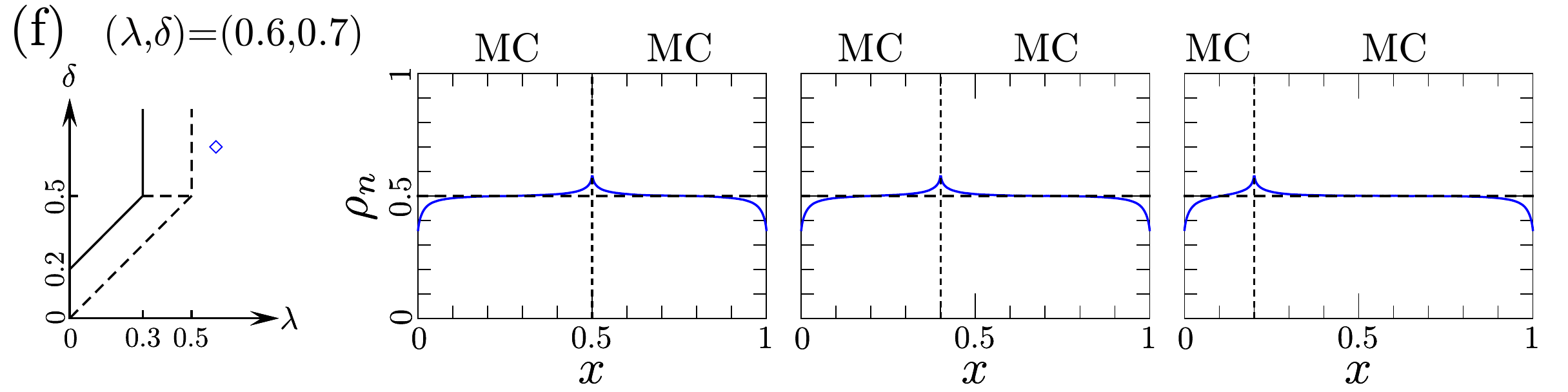}
\caption{Macroscopic density profiles for $ ( \gamma,\Omega ) = (0.3,0.4) $. In each row, a chosen set of $ (\lambda, \delta) $ is indicated, and its position is also displayed in the phase diagram. 
In each panel, the density profiles by simulations for $ L=1000 $ and $ \gamma'=0 $ \cite{bib:technical} (solid lines) 
are compared with the predictions $ \rho_n(x) $ (dashed lines),
and the vertical dotted line represents the MTOC position. 
For convenience, we wrote the types (LD, HD, MC, or a combination of them) of the realized density profiles 
above each panel. We expect that the deviations from the predictions are due to finite-size effects and boundary layers. 
\label{fig:macro}}
\end{center}
\end{figure}

Now we consider the condition \eqref{eq:condition} for the localization. 
In view of the relation \eqref{eq:relation}, Eq. \eqref{eq:condition} is satisfied when 
the both density profiles are of the LD type 
Otherwise $ \rho_1 ( 0 ) $ and $ \rho_2 ( 1) $ are independent from 
the MTOC position $c$. 
This is rephrased as 
\begin{align}\label{eq:localization-condition}
 \rho^\text{LD}_1 ( 0 ) < \min \{ 1 / 2 , 1-\delta \} ,
 \ \text{or equivalently } \rho^\text{LD}_2 ( 1 ) < \min \{ 1 / 2 , 1-\delta \} 
\end{align} 
for $ c= 1 /2 $. see Fig.~\ref{fig:macro} (a,d).
Under the conditions \eqref{eq:localization-condition},
 the macroscopic density profile of each MT is linear
in the interval $ 0 < x < c $ or $ c < x < 1 $, and 
if $c$ is slightly shifted from $1/2 $, we find 
\begin{align} 
 \rho_1 ( 0 ) = \lambda + c \Omega \lessgtr 
 \lambda + (1-c) \Omega = \rho_2 ( 1 ) \quad (\text{for } c \lessgtr 1 / 2 ) . 
\end{align} 
i.e. Eq. \eqref{eq:condition} is satisfied. The parameter region where the conditions~\eqref{eq:localization-condition} are satisfied is the localization phase given as Eq.~\eqref{eq:localization-phase}.

On the other hand, in the delocalization phase, the macroscopic density profiles near $x =0$ and $x=1$
become of the HD or MC type for $c=1/2$, see Fig.~\ref{fig:macro}~(b,c,e,f). 
Since \eqref{eq:HD} and \eqref{eq:MC} do not depend on the MTOC position, the relation 
$ \rho_1 ( 0 ) = \rho_2 ( 1 ) $ is kept in some interval $ u\le c \le v $ including $c = 1/2$: 
\begin{align}
 \rho_1 ( 0 ) < \rho_2 ( 1 ) \ (\text{for } c < u) , \quad 
 \rho_1 ( 0 ) = \rho_2 ( 1 ) \ (\text{for } u \le c \le v) ,\quad 
 \rho_1 ( 0 ) > \rho_2 ( 1 ) \ (\text{for } v < c ) .
\end{align}
(From the left-right symmetry, we have $ v=1- u $.)
The solution $c$ ($ 0 < c < 1/2 $) to 
\begin{align} \label{eq:=r10=min}
 \rho^\text{LD}_1 ( 0 ) = \min \{ 1 / 2 , 1-\delta \} 
\end{align} 
gives the minimum $u$, if it exits. 
In this case, $\rho^\text{LD}_1 ( x )$ ($ 0 < x < c $) is selected for $0 < c < u $,
see Fig.~\ref{fig:macro} (b,e). 
If there is no solution to \eqref{eq:=r10=min}, we have $u=0$,
and a LD is not realized near $ x=0 $ or $x=1$ even for very small $c$,
see Fig.~\ref{fig:macro} (c,f). 
Summarizing the results, we get the dashed lines in the phase diagram Fig.~\ref{fig:tip-density}.

Now we comment the trapping at the boundaries, in the subphase of `$ \sigma = \frac{1}{2\sqrt 3} $'. 
One naively thinks that we always have $ \rho_1 (0) =\rho_2 (1) $ and $ \rho^+_{L_1} =\rho^+_{L_2} $,
and expects the uniform distribution over $ 0<c<1 $. However, when $ L_1 \ll L_2 $ (i.e. $ c \approx 0 $), the finite-size effect on the MT $n=1$ is crucial, where the hydrodynamic description $ \rho_1(x) $ is no longer valid. We have not fully determined the form of $ \rho^+_{L_1} $, but there exists a parameter region where
 $\rho^+_{L_1} > \rho^+_{L_2} $ for $ L_1 \ll L_2 $, see e.g. $ \lambda = 0.6 $ of Fig.~5~(b). 
 This inequality for the tip densities induces trapping at the left boundary, since the system is often pushed leftward. 
As a result, the standard deviation becomes larger than the naive prediction $ \frac{1}{2\sqrt 3} $. A similar argument is possible for the trapping at the right boundary. 

We turn to the generalized model, where kinesin motors attach to the MTs with rates $ \Omega_\text{A1} $
and $ \Omega_\text{A2} $ for sites $ i< i^* $ and $i\ge i^*$, respectively. 
The attachment rate $ \Omega_\text{A1} $ is different from $ \Omega_\text{D} $
but we keep $ \Omega_\text{A2 } = \Omega_\text{D} =: \Omega $. 
The hydrodynamic equation for the MT $n=1$ is now 
\begin{align} 
\begin{cases}
 \partial_x [ \rho_1 (1- \rho_1 ) ] + \Omega_\text{A1} (1-\rho_1) - \Omega \rho_1 =0 & ( x < b ), \\
 (1-2\rho_1) ( \partial_x \rho_1 + \Omega ) = 0 & ( b < x < c ), 
\end{cases}
\end{align}
where $ b= i^*/L $. Solving this with the boundary conditions $ \lim_{ x\nearrow c } \rho_1(x) = \lambda $, 
and $ \lim_{ x\nearrow b } \rho_1(x) =\lim_{ x\searrow b } \rho_1(x) := \rho^* $,
we obtain 
\begin{align} 
\begin{cases}
 x = b
 - \frac{1}{ \Omega +\Omega_\text{A1} } 
 \big[ 2(\rho_1 - \rho^* ) + 
 \frac{ \Omega - \Omega_\text{A1} }{ \Omega +\Omega_\text{A1} } 
 \ln \frac{ \Omega_\text{A1} - ( \Omega + \Omega_\text{A1} )\rho_1 }{ 
 \Omega_\text{A1} - ( \Omega + \Omega_\text{A1} ) \rho^* } \big] & ( x < b ), \\
 \rho_1 (x) = \lambda - ( x - c ) \Omega& ( b < x < c ), 
\end{cases}
\end{align}
with $\rho^* = \lambda - (b-c) \Omega $. 
The density profile of the MT $n=2$ is unchanged. 
The average $ \langle c \rangle $ is given by solving 
$ \rho_{L_1}^+ =\rho_{L_2}^+ $, i.e. $ \rho_1(0)= \rho_2 (1) $, 
but one cannot get an explicit solution. Instead, one performs perturbative calculation around $ \epsilon =0 $ 
 with $ \Omega_\text{A1} = (1+\epsilon)\Omega $, leading to Eq.~\eqref{eq:first-order}.


\begin{thebibliography}{99}
\bibitem{bib:technical} 
Technically, in oder to measure the density profiles by simulations, one needs to consider conditional probabilities for a given MTOC position. Therefore, for the cases where the localization is realized with $ \gamma'>0$, it is difficult to obtain the density profiles for $c$ being far from $1/2$.
To avoid this difficulty, we used the model with $ \gamma' =0 $ for simulations.
\end{thebibliography}
\end{document}